\documentclass[prd, aps, twocolumn, showpacs,superscriptaddress]{revtex4-1}
\usepackage{amsbsy,amsmath,amssymb,amsthm,wasysym,amsfonts}
\usepackage{graphicx}
\usepackage{psfrag}
\usepackage{epstopdf}
\usepackage{color}
\usepackage{mathrsfs}
\usepackage{comment}
\usepackage[normalem]{ulem}

\DeclareSymbolFontAlphabet{\mathrsfs}{rsfs}
\DeclareMathAlphabet\mathbfcal{OMS}{cmsy}{b}{n}

\newcommand{\be}{\begin{equation}}  
\newcommand{\ee}{\end{equation}}
\newcommand{\bea}{\begin{eqnarray}}           
\newcommand{\eea}{\end{eqnarray}} 
\newcommand{\beqn}{\begin{eqnarray*}}
\newcommand{\eeqn}{\end{eqnarray*}}
\newcommand{\ba}{\begin{align}}
\newcommand{\ea}{\end{align}}

\def\lm{{\ell m}}   % This for \ell m

\def\ha{{\hat{a}}}
\def\ta{{\tilde{a}}}

\def\p4{{\psi_4}} % This is psi4
\usepackage{float}
\definecolor{cyan}{rgb}{0,0.9,0.9}
\definecolor{orange}{rgb}{0.9,0.5,0}
\definecolor{magenta}{rgb}{1,0,1}
\definecolor{purple}{rgb}{0.8,0.4,0.8}

\begin{document}

% =====================================
% titlepage
% =====================================
\title{iResum: a new paradigm for resumming gravitational wave amplitudes}

\author{Alessandro \surname{Nagar}}
\affiliation{Institut des Hautes Etudes Scientifiques, 91440
  Bures-sur-Yvette, France}
\affiliation{INFN, Sez. di Torino, Via P.~Giuria 1, 10125, Torino, Italy}
\author{Abhay \surname{Shah}}
\affiliation{Mathematical Sciences, University of Southampton, Southampton, SO17 1BJ, United Kingdom}

% --------------------------------------
\begin{abstract}
We introduce a new, resummed, analytical form of the post-Newtonian (PN), 
factorized, multipolar   amplitude    corrections  $f_{\lm}$ of the 
effective-one-body (EOB) gravitational waveform of spinning, 
nonprecessing, circularized, coalescing black hole binaries (BBHs). 
This stems from the following two-step paradigm: 
(i) the {\it factorization} of the orbital (spin-independent) terms in $f_{\lm}$; 
(ii) the {\it resummation} of the residual spin (or orbital) factors. We find that 
resumming the residual spin factor by taking its {\it inverse resummed} (iResum) 
is an efficient way to obtain amplitudes that are more accurate in the strong-field, 
fast-velocity regime. The performance of the method is illustrated on the  
$\ell=2$ and $m=(1,2)$ waveform multipoles, both for a test-mass orbiting 
around a Kerr black hole  and for comparable-mass BBHs. In the first case, 
the iResum $f_{\lm}$'s are much closer to the corresponding ``exact'' functions 
(obtained solving numerically the Teukolsky equation) up to the light-ring, 
than the nonresummed ones, especially when the black-hole spin is nearly 
extremal. The iResum paradigm is also more efficient than including higher 
post-Newtonian terms (up to 20PN order): the resummed 5PN information 
yields per se a rather good numerical/analytical agreement at the last-stable-orbit, 
and a well-controlled behavior up to the light-ring. 
For comparable mass binaries (including the highest PN-order information 
available, 3.5~PN), comparing EOB with Numerical Relativity (NR) 
data shows that the analytical/numerical fractional disagreement at merger, 
{\it without NR-calibration of the EOB waveform}, is generically reduced 
by iResum, from a $40\%$ of the usual  approach to just a few percents.
This suggests that EOBNR waveform models for coalescing BBHs may 
be improved using iResum amplitudes.
\end{abstract}
% --------------------------------------
\date{\today}


\maketitle

\section{Introduction}
Determining the physical properties of the binary black hole (BBH) mergers
GW150914~\cite{Abbott:2016blz} and GW151226~\cite{Abbott:2016nmj} required a large bank of (semi)-analytical 
gravitational wave (GW) templates~\cite{Taracchini:2013rva,Husa:2015iqa,Khan:2015jqa}.
The effective-one-body (EOB) theory~\cite{Taracchini:2013rva,Damour:2014sva,Szilagyi:2015rwa,Nagar:2015xqa} 
was essential to model gravitational waveforms from BBHs 
with total mass $M\equiv m_{1}+m_{2}$ larger than $4M_{\odot}$~\cite{TheLIGOScientific:2016wfe}.
One of the pillars of EOB theory is the factorized and resummed
(circularized) multipolar post-Newtonian (PN) waveform of  
Refs.~\cite{Damour:2007xr,Damour:2008gu} 
(generalized to spinning binaries in~\cite{Pan:2010hz}),
that radically improves the 1997 pioneering work of Ref.~\cite{Damour:1997ub} 
on PN fluxes resummation. The resummation makes this waveform better 
behaved in the strong-field, fast-velocity regime (i.e., up to merger), than 
the standard, Taylor-expanded, PN result~\cite{Blanchet:2013haa}
(the leading PN order being Einstein's quadrupole formula). 
The squared waveform multipoles, summed together, 
give the GW angular momentum flux emitted at infinity (or absorbed at the 
horizons~\cite{Nagar:2011aa,Taracchini:2013wfa,Damour:2014sva})
that provides the radiation reaction force driving the binary dynamics 
from the quasi-adiabatic circular inspiral through plunge and merger.
This paper proposes an additional factorization (and resummation) of the (residual) 
multipolar waveform amplitude correction for  nonprecessing, spinning,Ê~BBHs of Refs.~\cite{Pan:2010hz,Taracchini:2012ig} 
to improve its behavior close to merger, both for large-mass-ratio 
and comparable-mass-ratio binaries, thus helping the development of 
EOB-based waveform models~\cite{Taracchini:2013rva,Nagar:2015xqa}.
We mostly use units with $c=G=1$.
\section{Results: the large-mass-ratio limit.} The factorized multipolar waveform 
for circularized, nonprecessing, BBHs with total mass $M$ and 
spins $S_1$ and $S_2$  reads (see e.g. Eq.(75)-(78) of~\cite{Damour:2014sva})
\be
\label{eq:hlm}
h_\lm(x)=h_\lm^{(N,\epsilon)}\hat{h}^{\rm tail}_\lm \hat{S}^{(\epsilon)}_{\rm eff}f_\lm(x,S_1,S_2),
\ee
where $x=(G M\Omega/c^{3})^{2/3}={\cal O}(c^{-2})$ is the PN-ordering frequency parameter
[we recall that $n$-PN order means $O(c^{-2n})$ in the equations of motion], 
$h_\lm^{(N,\epsilon)}$ the Newtonian prefactor, where $\epsilon=0,1$ is 
the parity of the considered multipole, i.e. of $\ell+m$ (see also Eq.~(78) of~\cite{Damour:2014sva}); 
$\hat{h}^{\rm tail}_\lm\equiv T_\lm e^{{\rm i}\delta_\lm}$  is the (complex) factor that accounts 
for the effect of tails~\cite{Damour:2007xr,Damour:2008gu}; the third factor, $\hat{S}_{\rm eff}^{(\epsilon)}$ 
is the usual parity-dependent source term defined as the effective EOB Hamiltonian when $\epsilon=0$
or the Newton-normalized orbital angular momentum when $\epsilon=1$~\cite{Damour:2008gu}.
The fourth factor, $f_\lm$, is the residual amplitude correction; its 
further resummation when $\nu\equiv m_1 m_2/M^2\neq 0$ and $S_{1,2}\neq 0$, 
depends on the parity of $m$~\cite{Nagar:2015xqa,Taracchini:2013rva}. 
When $m$ is even, it is resummed as $f_\lm=(\rho_\lm)^\ell$ with $\rho_\lm$ 
given as 
\be
\label{eq:rho_additive}
\rho_\lm = \rho_\lm^{\rm orb}+\rho_\lm^{\rm S},
\ee
where  $\rho_\lm^{\rm orb}$ is the {\it orbital} (spin-independent) contribution 
and $\rho_\lm^{\rm S}$ is the spin-dependent part. Both functions are given as 
PN-expansions. The PN-accuracy of $\rho_\lm$'s currently used in 
EOB models~\cite{Nagar:2015xqa,Taracchini:2013rva} is relatively low:
$\rho_\lm^{\rm orb}$ is taken at $3^{+2}$~PN accuracy~\cite{Damour:2009kr}
(i.e., up to  3PN with complete mass-ratio dependence plus 4PN and 5PN test-particle, $\nu=0$, terms)
while $\rho_\lm^{\rm S}$, e.g., as used in the 
{\tt SEOB\_ihes} model~\cite{Damour:2014sva,Nagar:2015xqa}, 
has leading-order (LO, i.e. 2PN) spin-spin ($\text{S}^{2}$) terms 
and up to next-to-leading-order (NLO, i.e. 2.5PN) spin-orbit (SO) 
terms.
Thanks to recent analytical work~\cite{Marsat:2013wwa,Marsat:2014xea,Bohe:2015ana}, 
the $f_{\lm}$'s (for both even and odd $m$'s) can be obtained 
up to next-to-next-to-leading-order (NNLO) in SO, NLO 
in ${\rm S^2}$ and LO in ${\rm S^3}$.
In the large-mass-ratio (i.e., test-particle) case ($m_1\gg m_2$, i.e. $\nu=0$), 
$\rho_\lm^{\rm orb}$ is known at 22.5PN~\cite{Fujita:2014eta}; 
for a spinning BH, the fluxes were obtained at 20PN~\cite{Shah:2014tka} 
extracting the PN coefficients from numerical data (see also 
Ref.~\cite{Fujita:2014eta} for a fully analytical calculation at 11PN).

\begin{figure}[t]
\center
\includegraphics[width=0.40\textwidth]{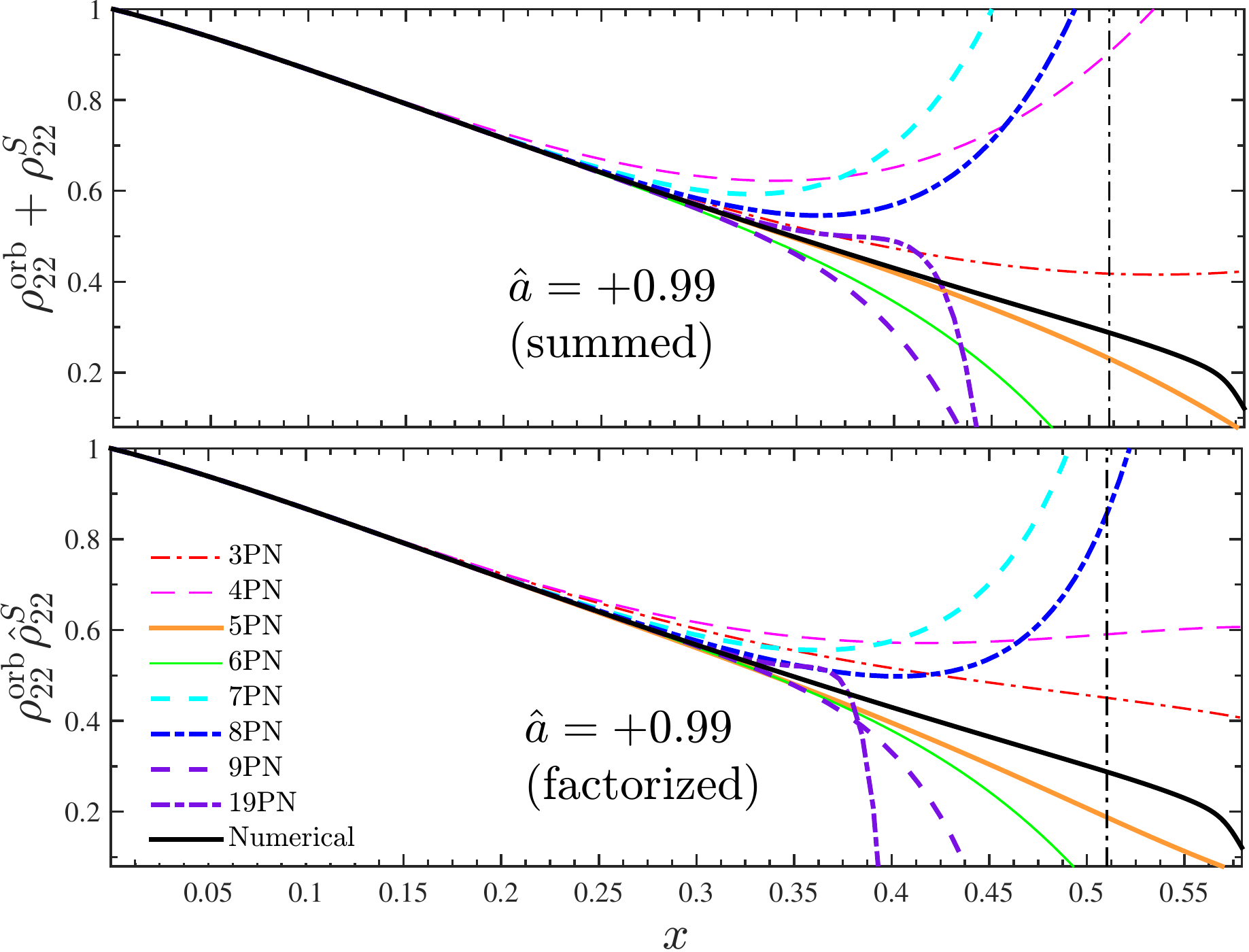}
\caption{\label{fig:fig1}Large-mass-ratio limit, BH spin $\ha=+0.99$: 
various PN approximants of $\rho_{22}$ with additive 
spin-dependence (top, Eq.~\eqref{eq:rho_additive}) or factorized 
(bottom, Eq.~\eqref{eq:rhoTT}). Increasing the PN order does 
not reduce the (large) disagreement with the numerical 
(``exact'') curve towards the LSO (vertical line). 
The $x$-axis ends at the light-ring frequency.}
\end{figure}
%------------------

Let us first focus on the (nonspinning) test-particle case, $m_2\ll m_1$, 
around a spinning BH. We shall use dimensionless
spin variables $\ha_{1,2}\equiv S_{1,2}/(m_{1,2})^2$, which in the 
non-spinning test-particle case yield $\ha_2=0$ and $\ha\equiv \ha_1$. 
We discuss here explicitly only the $\ell=m=2$ 
and $\ell=2$, $m=1$ modes, postponing elsewhere 
the investigation of other multipoles. From the 20PN flux
of Ref.~\cite{Shah:2014tka}, we obtain the 20PN-accurate
$\rho_{22}$, as in Eq.~\eqref{eq:rho_additive}, 
and explore its behavior for the most demanding case $\ha=+0.99$
(our conclusions actually hold for any smaller value of $\ha$).
Figure~\ref{fig:fig1}, top, contrasts various PN-approximants $\rho_{22}$
with the  ``exact'' (numerical) curve $\rho_{22}^{\rm Num}$ (black) stopping 
at the light-ring. This curve was factorized from the energy fluxes 
computed by S.~Hughes solving numerically the Teukolsky 
equation with high accuracy (fractional uncertainty~$10^{-14}$)~\cite{Taracchini:2013wfa}.
The vertical dash-dotted line marks  the last-stable-orbit (LSO). 
The overall bad behavior of the various PN-approximants is evident, with the 
19PN one (as an example of high-PN information) not showing 
any improvement with respect to any lower-PN curve. The strong-field
inaccuracy of the PN-expanded $\rho_{22}$ when $\ha\gtrsim 0.7$ 
was already noted (at 4PN only) in~\cite{Pan:2010hz}  
and improvements were proposed 
(see also~\cite{Fujita:2014eta,Isoyama:2012bx,Johnson-McDaniel:2014lia}),
though fine-tuned to the test-particle limit.

\begin{figure}[t]
\center
\includegraphics[width=0.42\textwidth]{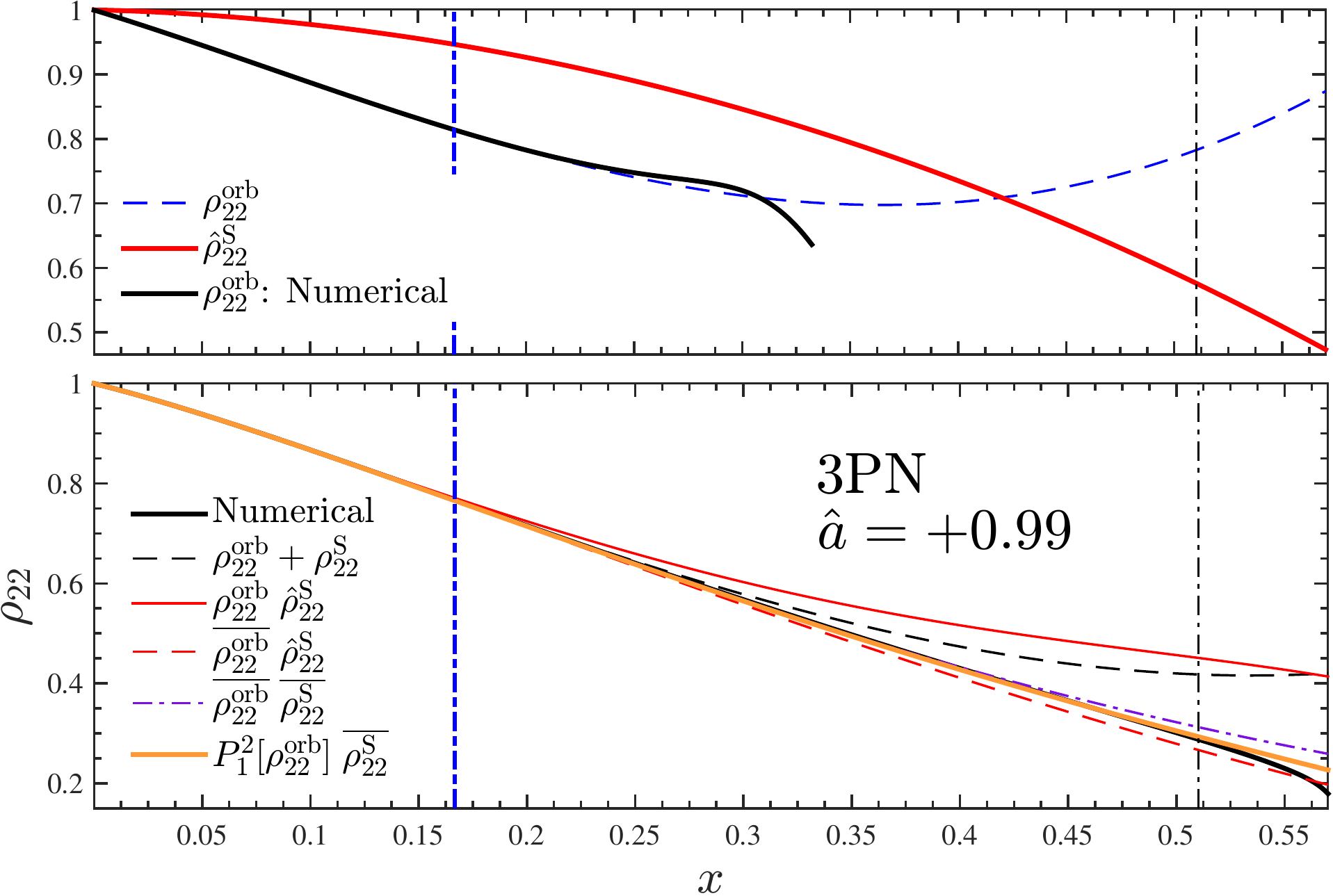}
\caption{\label{fig:fig3PN}Large-mass-ratio limit, 3PN accuracy. Top: the
behavior of the two factors $\rho_{22}^{\rm orb}$ and $\hat{\rho}_{22}^{\rm S}$.
The zero-spin (orbital), exact, curve (black, ending at the Schwarzschild light-ring, $x=1/3$) 
is included for comparison. Bottom: comparing nonresummed and resummed functions. 
Note the improvement towards LSO (and beyond) due to resummations.}
\end{figure}

To devise a new form of $\rho_{22}$ that is well-behaved
up to the light-ring and valid also beyond the test-particle case, 
we: (i)  {\it factorize} the orbital part and (ii)  {\it resum} the 
resulting spin (or orbital) factors. At $n$-PN order 
the {\it orbital factorized} amplitudes read
\be
\label{eq:rhoTT}
\tilde{\rho}_\lm(x;\,\hat{a}) \equiv \rho_\lm^{\rm orb}\hat{\rho}_\lm^{\rm S},
\ee
where $\hat{\rho}_\lm^{\rm S}\equiv T_n\left[1+\rho_\lm^{\rm S}(x;\,\hat{a})/\rho_\lm^{\rm orb}(x)\right]$, with $T_n[\dots]$ 
denoting the Taylor expansion at order $n$.  The 5PN $\hat{\rho}_{22}^{\rm S}$ reads
\begin{align}
\label{eq:trho22S}
\hat{\rho}_{22}^{\rm S_{\rm 5PN}}& =1-\dfrac{2}{3}\ha x^{3/2}+\dfrac{\ha^2}{2} x^2 -\dfrac{145}{63}\ha x^{5/2}
                    +\dfrac{109}{126}\ha^2 x^3 \nonumber\\
                   & +\left(-\dfrac{9808}{3969}\ha + \dfrac{\ha^3}{3}\right)x^{7/2}
                    +\left(\dfrac{7207}{2646}\ha^2 - \dfrac{\ha^4}{8}\right)x^4\nonumber\\
                  & +\left[\dfrac{925}{1134}\ha^3+\ha\left(-\dfrac{5094410}{305613}+\dfrac{16}{3}\text{eulerlog}_2(x)\right)\right]x^{9/2} \nonumber\\
                  & + \left(\dfrac{1753858}{305613}\ha^2-\dfrac{59}{252}\ha^4\right)x^5,
\end{align}
where $\text{eulerlog}_2(x)=\gamma+2\log2+1/2\log x$; the 5PN $\rho^{\rm orb}_{22}$ 
is given in Eq.~(50) of~\cite{Damour:2008gu}. To understand how
to proceed further, let us take the 3PN-truncations of $(\rho_{22}^{\rm orb},\rho_{22}^{\rm S})$ 
for $\ha=+0.99$, see Fig.~\ref{fig:fig3PN}, top. The bottom panel
contrasts the exact curve with various ways of using this 3PN information.

\begin{figure}[t]
\center
\includegraphics[width=0.40\textwidth]{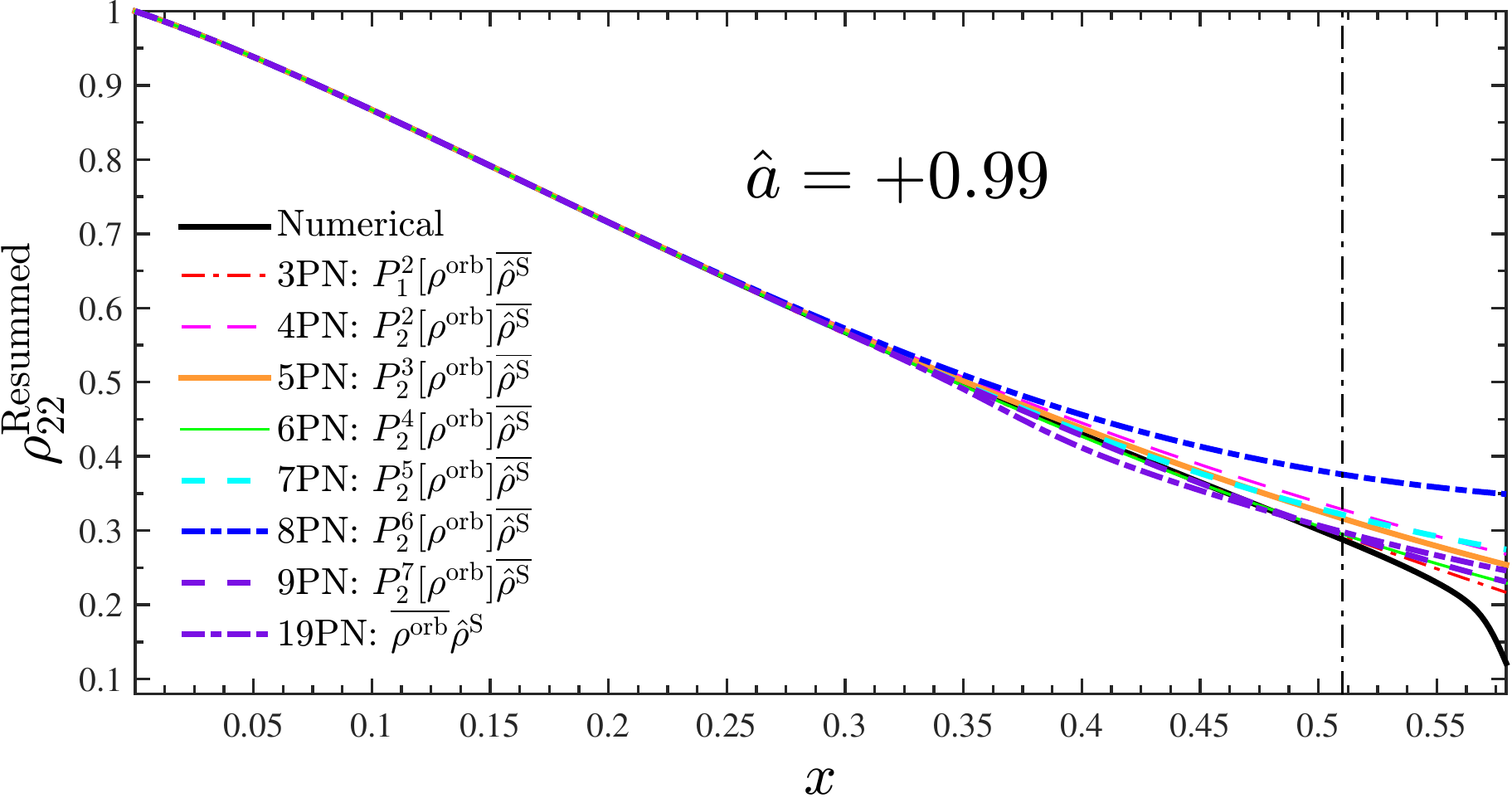}
\caption{\label{fig:resum}Resumming separately $\rho^{\rm orb}_{22}$ and 
$\hat{\rho}^{\rm S}_{22}$: the reduction of the PN ambiguity towards the
LSO (vertical line) as well as the closeness to the numerical
curve is remarkable (and consistent) for any PN order 
(contrast with Fig.~\ref{fig:fig1}).}
\end{figure}
%------------------
Inspecting Fig.~\ref{fig:fig3PN}, one sees that up to, say, the Schwarzschild LSO 
frequency, $x_{\rm LSO}^{\rm Schw}=1/6=0.1\bar{6}$ (thick, blue, vertical line), 
$\rho_{22}^{\rm orb}$ and $\rho_{22}^{\rm S}$ somehow compensate each other, 
so that their product $\rho_{22}^{\rm orb}\hat{\rho}^{\rm S}_{22}$ is on top 
of $\rho_{22}^{\rm Num}$. 
For larger values of $x$, the orbital factor takes over to determine a large difference 
at $x_{\rm LSO}^{\rm +0.99}$ (thinner vertical line in Fig.~\ref{fig:fig3PN}). 
The behavior of the 3PN-accurate $\rho_{22}^{\rm orb}$ and $\hat{\rho}_{22}^{\rm S}$ 
remains essentially unchanged also at higher PN orders (see Fig.~\ref{fig:fig1}, bottom), 
with the orbital factor dominating when the $\tilde{\rho}_{22}$'s increase towards the LSO, 
and the spin factor dominating  when they decrease and get negative~\footnote{An exception is represented by 
the 7PN and 8PN approximants, since the corresponding $\hat{\rho}^{\rm S}_{22}$'s change 
curvature around the LSO crossing frequency so that the two effects accumulate}. 
Both $(\rho_{22}^{\rm orb}$ and $\hat{\rho}_{22}^{\rm S})$ can be separately 
improved by replacing them with suitable resummed expressions, so that their 
growth (or decrease) is milder towards $x_{\rm LSO}^{\rm +0.99}$; this eventually
reduces the global disagreement of their product with $\rho_{22}^{\rm Num}$.
There are several ways of doing so. A simple approach is to resum 
one of the two factors (or both) by taking its inverse, i.e. replacing 
a function $f(x)$ with its inverse resummed (the ``iResum'') representation 
\be
\label{eq:iResum}
\overline{f(x)}\equiv \left(T_n\left[(f(x))^{-1}\right]\right)^{-1}.
\ee
Figure~\ref{fig:fig3PN}, bottom, illustrates the remarkable efficiency
of this procedure when it is applied just to $\rho_{22}^{\rm orb}$ (red-dashed 
line) or to both $\rho_{22}^{\rm orb}$ and $\hat{\rho}_{22}^{\rm S}$ (purple, dash-dotted line). 
An even better approximant even beyond the LSO is obtained 
by taking the (2,1) Pad\'e approximant of $\rho_{22}^{\rm orb}$ 
multiplied by $\overline{\hat{\rho}_{22}^{\rm S}}$.  It is remarkable that 
the 3PN information, once properly resummed, can yield accurate 
predictions ($1\%$ fractional difference) for such an extremal value of $\ha$
\footnote{Note however that it becomes less accurate for smaller values of $\ha$, 
because $\rho_{22}^{\rm orb}$ is only 3PN.} .
The method consistently works up to 9PN, so that the large
PN-ambiguity of Fig.~\ref{fig:fig1} is dramatically reduced, see Fig.~\ref{fig:resum}.
For $n$-PN order, we always take the $(n-2,2)$ Pad\'e approximant of 
$\rho_{22}^{\rm orb}$, since it is the only one without spurious poles.
We also note that, up to 5PN, also functions of the form
$\overline{\rho^{\rm orb}_{22}}\hat{\rho}_{22}^{\rm S}$ are rather accurate
up to $x_{\rm LSO}^{+0.99}$. Unfortunately, the same is not true
up to 9PN because of a pole in $\overline{\rho_{22}^{\rm orb}}$ 
that progressively moves within the $x$-domain of interest; it is however 
possible to take $\rho_{\rm 22}^{\rm orb}\overline{\hat{\rho}_{22}^{\rm S}}$ 
instead, although globally its performance is slightly worse than  
that of Fig.~\ref{fig:resum} (as for 7PN and 8PN).
For higher PN  ($10\leq n\leq 20$) the simple recipes discussed 
above yield, on average, less good and less robust results. 
The only exception is given by the 19PN, that  can be consistently resummed as
$\overline{\rho^{\rm orb}_{22}}\hat{\rho}_{22}^{\rm S}$ (see Fig.~\ref{fig:resum}).
Summarizing, any PN-resummed function of Fig.~\ref{fig:resum} 
can be used to construct reliable radiation reaction for BBH coalescence 
in the test-particle limit~\cite{Taracchini:2014zpa,Harms:2014dqa} so to 
improve several results (see e.g.~\cite{Harms:2014dqa,Nagar:2014kha}) 
that are biassed by its poor accuracy for high spins~\footnote{Possibly, without 
the need of fitting higher-order effective terms from the numerical 
data as in~\cite{Nagar:2011aa,Taracchini:2013wfa}.}.

\begin{figure}[t]
\center
\includegraphics[width=0.40\textwidth]{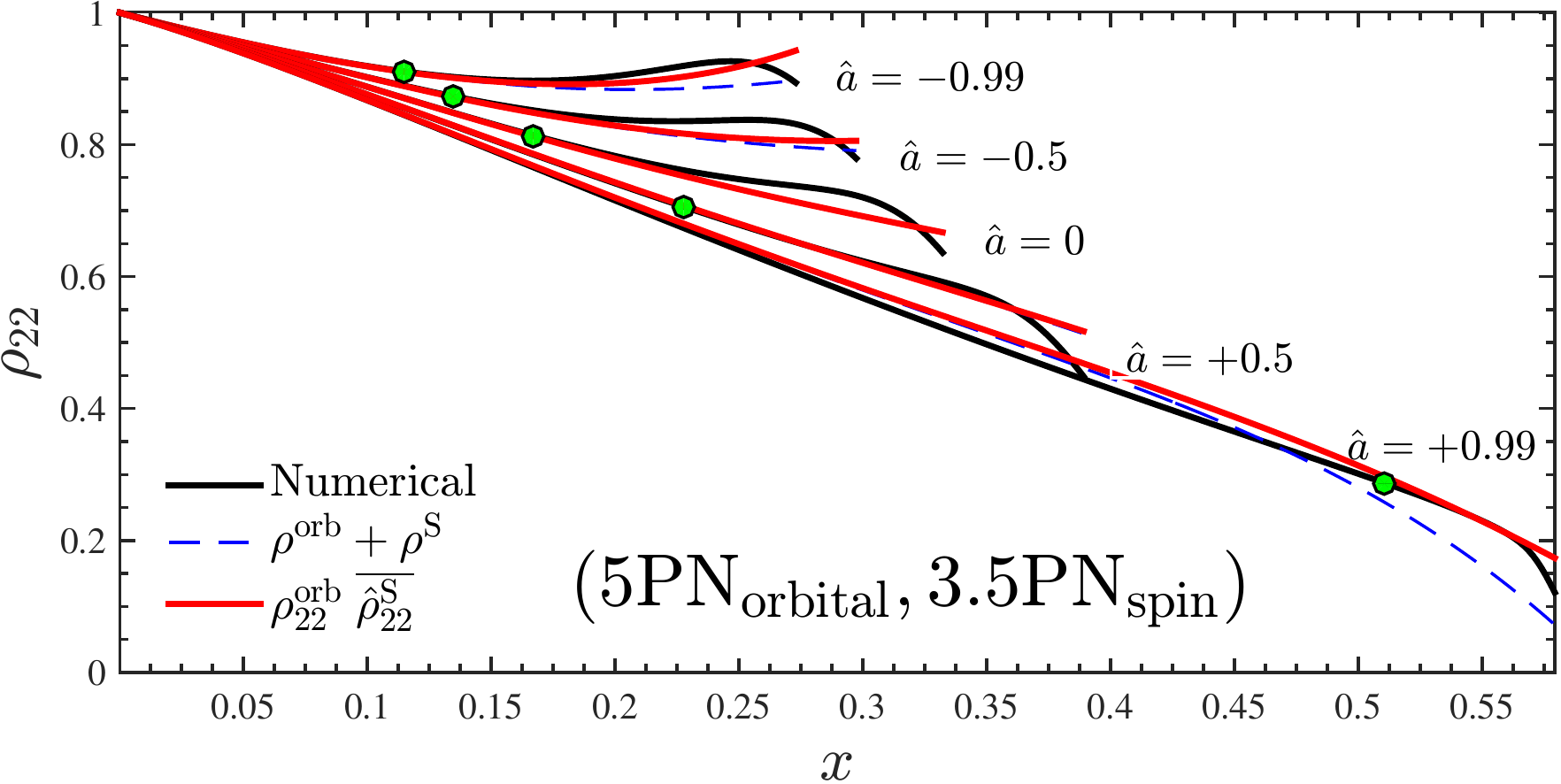}
\caption{\label{fig:fig2}Large mass-ratio limit: $\rho_{22}^{\rm orb}$ at 5PN and $\rho_{22}^{\rm S}$ at 3.5PN.
Resumming  only the spin-dependent factor yields a remarkable agreement 
with the numerical curve through LSO (and beyond), notably when $\ha\approx 1$.}
\end{figure}

To prepare the ground for the $\nu\neq 0$ case, where $\rho_{22}^{\rm orb}$
is taken at $3^{+2}$~PN and $\rho_{22}^{S}$ at 3.5PN 
(i.e., including NNLO SO, NLO in ${\rm S^2}$  and LO \text{$S^{3}$}, see Eq.~\eqref{eq:trho22S})
we want to explore the efficiency of iResum also with this particular PN truncation.
Although the above analyses suggest to resum both $\rho^{\rm orb}_{\lm}$ 
and $\hat{\rho}_{\lm}^{\rm S}$ also when $\nu\neq 0$ 
for consistency, here we do it only on $\hat{\rho}^{\rm S}_{\lm}$, keeping $\rho_{\lm}^{\rm orb}$ 
nonresummed, as it is in well-established, NR-calibrated, EOBNR nonspinning 
models~\cite{Nagar:2015xqa}. We take then $\rho^{\rm orb}_{22}$ at 5PN 
and $\hat{\rho}^{\rm S}_{22}$ at 3.5PN and replace $\hat{\rho}^{\rm S}_{22}$
with $\overline{\hat{\rho}^{\rm S}_{22}}$. Figure~\ref{fig:fig2} illustrates the quality of 
this choice for a few values  of $\ha\in [-0.99,0.99]$. The plot also exhibits 
the additive $\rho_{22}^{\rm 5PN+3.5PN}$ of  Eq.~\eqref{eq:rho_additive}. 
The filled circles indicate LSO frequencies. The numerical-analytical 
fractional difference is at most of $6\%$ before the LSO for $\ha=+0.99$, and
always smaller for other values of $\ha$ (we also checked the agreement for
other intermediate values of $\ha$, not shown in the plot). The fractional difference 
at LSO is $3\%$ for $\ha=+0.99$, $1\%$ for $\ha=0.7$ and $\leq 0.2\%$ when $\ha\leq 0.5$.

\begin{figure}[t]
\center
\includegraphics[width=0.42\textwidth]{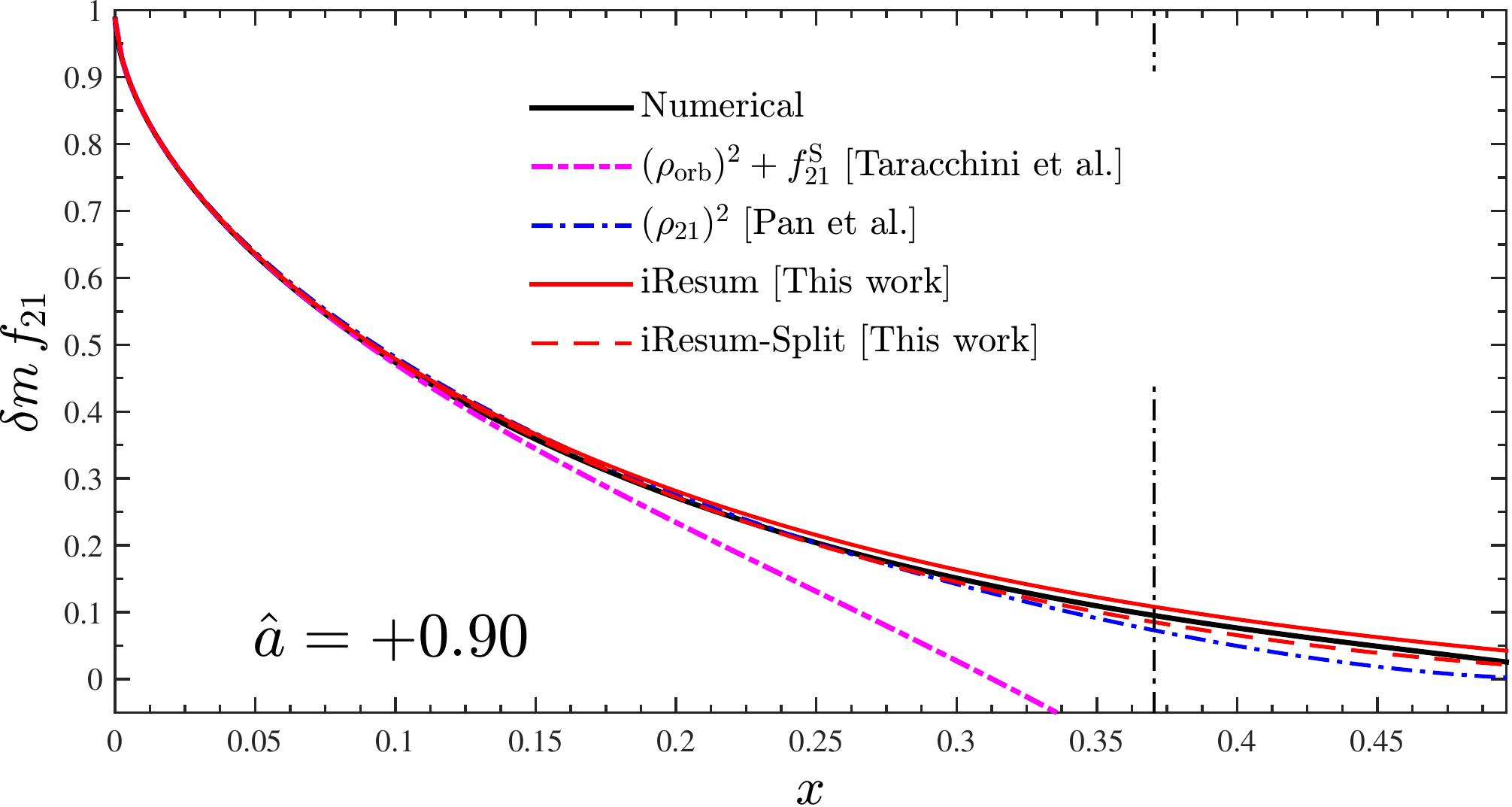}
\caption{\label{fig:fig3}Large mass-ratio limit, $\ha=+0.9$. 
Contrasting various representation of $\delta mf_{21}$, with
5PN-accurate $\rho_{21}^{\rm orb}$ and 2.5PN-accurate $\tilde{f}_{21}^{\rm S}$. 
The $x$-axis ends at the light-ring and the vertical line
marks the LSO. The defactorized amplitude of~\cite{Taracchini:2012ig,Damour:2014sva}, 
Eq.~\eqref{eq:dflm},  is largely inaccurate; 
by contrast, any of its resummed versions (obtained taking the 
inverse of spin-dependent parts) remains rather close to the exact one 
up to the light-ring.}
\end{figure}

Implementing a similar procedure when $m$ is odd calls for some distinguos.
Following Refs.~\cite{Taracchini:2012ig,Damour:2014sva}, 
the straightforward computation of $\rho_\lm$ as when $m$ is even, 
though doable in the test-particle limit, is unfit to the equal-mass case, 
because of formally singular terms. The solution~\cite{Taracchini:2012ig,Damour:2014sva}, 
is to {\it defactorize} the equal-mass vanishing 
factor, $X_{12}\equiv X_1-X_2=\delta m=(m_1-m_2)/M$,
from  $h_{\lm}^{\rm N}$ and use instead 
\be
\label{eq:dflm}
\delta m f_\lm(x,S_1,S_2)=X_{12} \left(\rho_\lm^{\rm orb}\right)^\ell+\tilde{f}_\lm^{\rm S}
\ee
where the $X_{12}$-rescaled functions $\tilde{f}_\lm^{\rm S}$, 
are given, at NLO, in Eqs.~(90)-(94) of~\cite{Damour:2014sva}. 
In the test-particle limit ($\nu=0$ and $X_{12}=1$), 
the $\ha$-dependence up to 3.5PN is given in Eq.~(28b) of~\cite{Pan:2010hz}.
Following the same rationale behind choosing 5PN+3.5PN 
for the $(2,2)$ mode, we take $\rho_{21}^{\rm orb}$ 
at 5PN and $\tilde{f}_{21}^{\rm S}$ at 2.5PN, because 
this is the highest-PN spin term known with $\nu\neq 0$ (see below).
Figure~\ref{fig:fig3} refers to $\ha=+0.9$, and illustrates 
the large disagreement between $\delta m f_{21}$ of 
Eq.~\eqref{eq:dflm} and $\delta m f_{21}^{\rm Num}$ 
built up to the LSO (vertical line). 
One can improve $\delta m f_{21}$ as in the
$m$-even case: (i) by factoring  
$(\rho_{\lm}^{\rm orb})^\ell$ and defining
\be 
\delta m\tilde{f}_\lm=(\rho_{\ell m}^{\rm orb})^\ell\hat{f}_{\lm}^{\rm S},
\ee
where $\hat{f}_{\lm}^{\rm S}\equiv T_{N}\left[X_{12}+\tilde{f}_{\lm}^{\rm S}/(\rho_{\lm}^{\rm orb})^\ell\right]$,
and (ii) by resumming this latter taking its inverse, Eq.~\eqref{eq:iResum}.
The result (red, solid, line in Fig.~\ref{fig:fig3}) remains very 
close to the numerical curve up to light-ring.
Figure~\ref{fig:fig3} also illustrates (blue line, dash-dotted) 
the performance of the ``standard'' $f_{21}=(\rho_{21})^2$ as given 
by Eq.~(29b) of Pan et el.~\cite{Pan:2010hz}, without any additional 
resummation~\footnote{Actually, $\rho_{21}$ can be further improved by doing the orbital
factorization and the straight iResum as when $m$ is even. Though this is fine-tuned to 
the test-particle case, it can be useful for building radiation reaction to drive the transition 
from inspiral to plunge into a quasi-extremal Kerr black 
hole~\cite{Taracchini:2014zpa,Nagar:2014kha,Harms:2015ixa,Gralla:2016qfw}.}.

\begin{figure}[t]
\center
\includegraphics[width=0.45\textwidth]{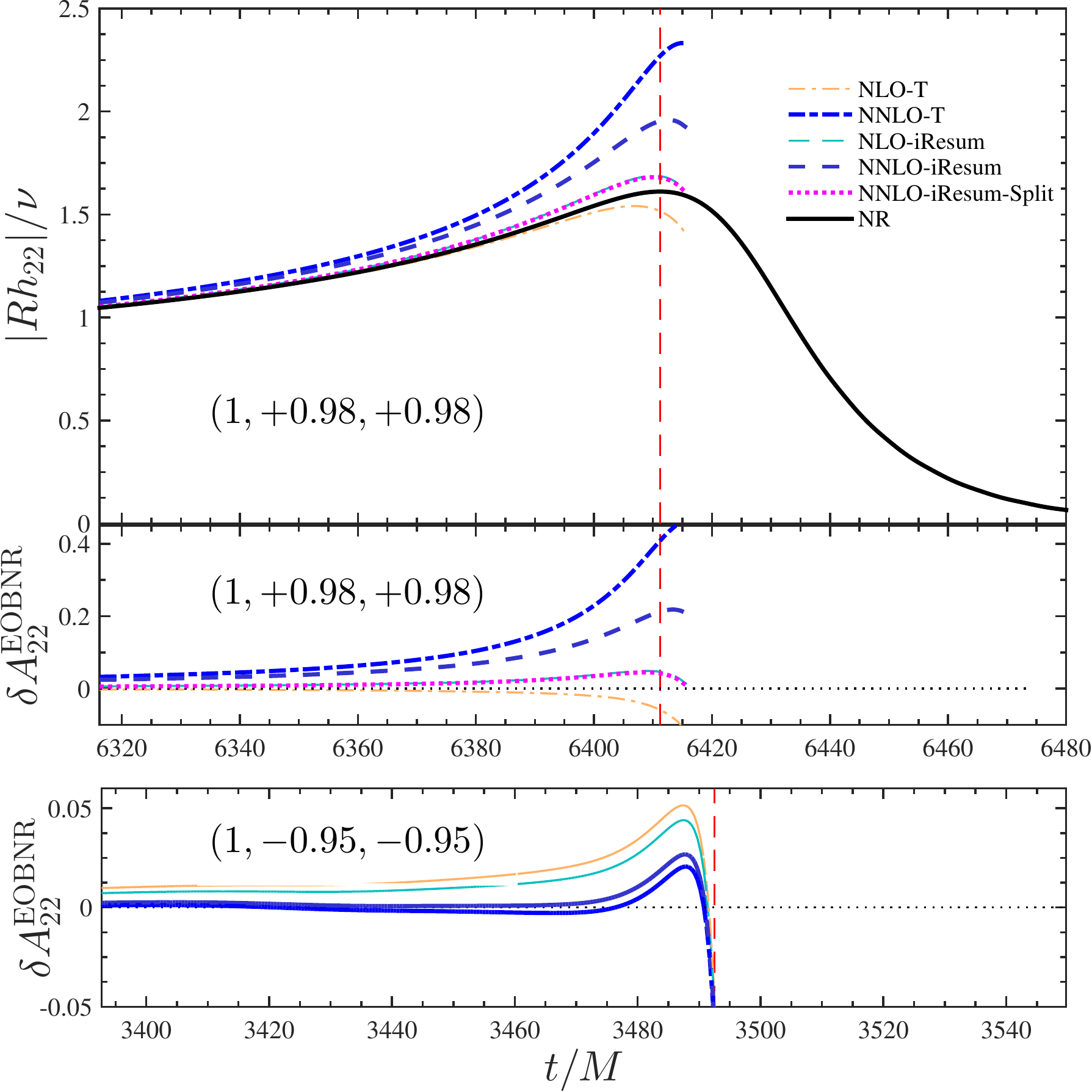}
\caption{\label{fig:A22} Equal-mass, equal-spin case, EOB/NR comparison: the
spread between NNLO and NLO is smaller for iResum than for the 
Taylor-expanded (T) residual amplitudes. The iResum waveform is always 
closer to the NR one at merger (red vertical line).}
\end{figure}
%-----------------------------
%------------------
\section{Results: the $\nu\neq 0$ case.}
\begin{figure}[t]
\center
\includegraphics[width=0.45\textwidth]{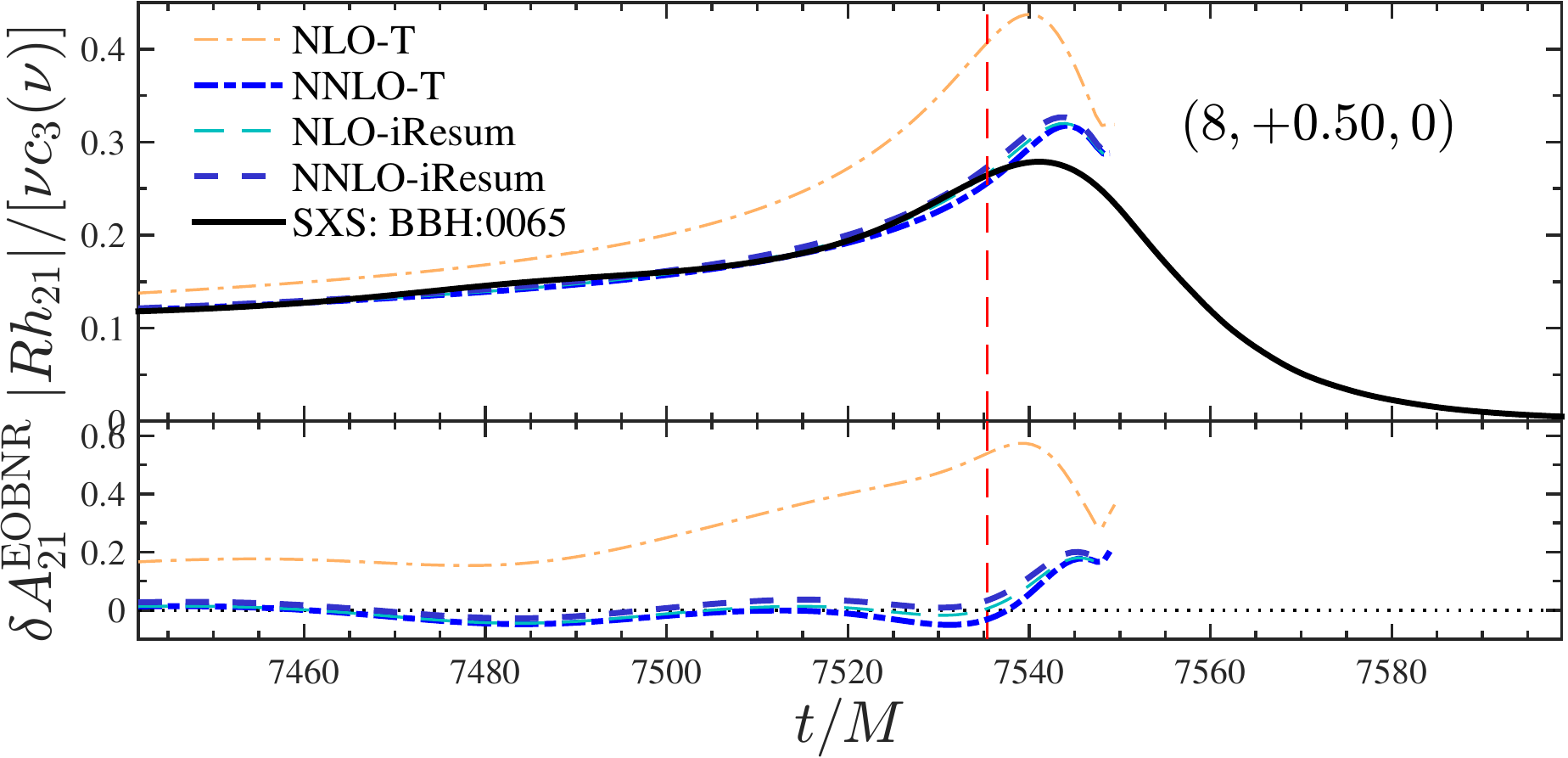}
\caption{\label{fig:A21} Same NNLO/NLO/NR comparison as Fig.~\ref{fig:A22}, but for the (2,1) 
mode, mass ratio $m_{1}/m_{2}=8$ and spins $(\ha_{1},\ha_{2})=(+0.5,0)$. The efficiency
of iResum on NLO is remarkable.}
\end{figure}

We obtained the $\nu$-dependent  $\hat{\rho}_{22}^{\rm S}$ and $\hat{f}_{21}^{\rm S}$ 
at the highest present PN accuracy from the multipolar decomposition  of 
the total flux (given to us by S.~Marsat and A.~Bohe from their summed 
result~\cite{Bohe:2015ana,Marsat:2013wwa}) that is known at NNLO, NLO and LO for SO,  
${\rm S^{2}}$ and ${\rm S^{3}}$ terms respectively. Expressions are simplified using
$\ta_{1,2}\equiv X_{1,2}\ha_{1,2}= S_{1,2}/(M m_{1,2})$, 
$\ha_{0}\equiv \ta_{1}+\ta_{2}$ and $\tilde{a}_{12}\equiv\ta_{1}-\ta_{2}$, so to get
\begin{align}
\label{eq:trho22}
&\hat{\rho}_{22}^{\rm S}=1-\left[\dfrac{\ha_{0}}{2}+\dfrac{1}{6}X_{12}\ta_{12}\right]x^{3/2}+\dfrac{1}{2}\ha_{0}^{2}x^{2}\nonumber\\
&+\Bigg[\left(-\dfrac{337}{252}+\dfrac{73}{252}\nu\right)\ha_{0}-X_{12}\ta_{12}\left(\dfrac{27}{28}+\dfrac{11}{36}\nu\right)\Bigg]x^{5/2}\nonumber\\
&+\Bigg[\dfrac{221}{252}X_{12}\ta_{12}\ha_{0}-\ha_{0}^{2}\left(\dfrac{1}{84}+\dfrac{31}{252}\nu\right)\nonumber\\
&\times\left(1-\dfrac{\ta_{1}\ta_{2}}{\ha_{0}^{2}}\dfrac{264-103\nu}{3+31\nu}\right)\Bigg]x^{3}
%&-(\ta_{1}^{2}+\ta_{2}^{2})\left(\dfrac{1}{84}+\dfrac{31}{252}\nu\right)+\ta_{1}\ta_{2}\left(\dfrac{43}{42}-\dfrac{55}{84}\nu\right)\Bigg]x^{3}\nonumber\\
+\Bigg[\ha_{0}\Bigg(-\dfrac{2083}{2646}\nonumber\\
&+\dfrac{123541}{10584}\nu+\dfrac{4717}{2646}\nu^{2}\Bigg)+X_{12}\ta_{12}\Bigg(-\dfrac{13367}{7938}\nonumber\\
&+\dfrac{22403}{15876}\nu+\dfrac{25}{324}\nu^{2}\Bigg)+\dfrac{7}{12}\ha_{0}^{3}-\dfrac{1}{4}X_{12}\ta_{12}\ha_{0}^{2}\Bigg]x^{7/2},
\end{align}
and 
\be
\label{eq:hatfS}
\hat{f}^{\rm S}_{21}=X_{12}f_{0}^{\rm S}-\dfrac{3}{2}\ta_{12}x^{1/2}f^{\rm S}_{1},
\ee
where
\begin{align}
\label{eq:f0S}
f_{0}^{\rm S}&=1-\dfrac{13}{84}\ha_{0}x^{3/2}+\dfrac{3}{8}\ha_{0}^{2}\left(1+\dfrac{4}{3}\dfrac{\ta_{1}\ta_{2}}{\ha_{0}^{2}}\right)x^{2}\nonumber\\
                    &+\ha_{0}\left(-\dfrac{14705}{7056}+\dfrac{12743}{7056}\nu\right)x^{5/2},\\
\label{eq:f1S}
f_{1}^{\rm S}&=1-\left(\dfrac{349}{252}+\dfrac{74}{63}\nu\right)x\nonumber\\
                    & +\dfrac{9}{4}\ha_{0}x^{3/2}-\left(\dfrac{3379}{21168}-\dfrac{4609}{10584}\nu+\dfrac{39}{392}\nu^{2}\right)x^{2},
\end{align}
and the $S^{3}$ term is here omitted for simplicity.
Figures~\ref{fig:A22} and~\ref{fig:A21} illustrate the benefits of iResum when 
applied to (truncations of) the above expressions. The figures compare
EOB waveform amplitudes to NR waveform amplitudes from the 
SXS catalog~\cite{SXS:catalog} for a few meaningful choices of mass ratio and spins.
For all EOB waveforms, the underlying EOB dynamics is that of the 
{\tt SEOBNR\_ihes} model of Ref.~\cite{Nagar:2015xqa}, with the NR-calibration 
of the parameters $(a_{6}^{c}(\nu),c_{3}(\ta_{1},\ta_{2},\nu)$ provided by Eqs.~(5) 
and (11) therein. By contrast, the waveform is {\it purely analytical} without 
the NR-calibrated next-to-quasi-circular (NQC) factors and ringdown (see~\cite{Nagar:2015xqa} 
for the performance of the full EOB waveform), that is why it stops just after 
merger (dashed vertical line).
The figures also include EOB-NR fractional differences, i.e.
$\delta A_{\lm}^{\rm EOBNR}\equiv (A^{\rm EOB}_{\lm}-A_{\lm}^{\rm NR})/A_{\lm}^{\rm NR}$,
where $A_{\lm}\equiv |Rh_{\lm}|$. For $(\ell,m)=(2,2)$ the waves are aligned 
using a standard procedure~\cite{Nagar:2015xqa} during the early inspiral; 
for (2,1) around merger, so to better highlight the differences there. 
With the label NLO we actually  indicate the spin information 
included in~\cite{Nagar:2015xqa}, i.e. NLO in the 
SO sector and LO {\it only} in SS, while NNLO refers to the complete 
Eqs.~\eqref{eq:trho22}-\eqref{eq:f1S} above.
We obtained $\overline{\hat{\rho}_{22}^{\rm S}}$ likewise the $\nu=0$ case. 
To iResum $\hat{f}^{\rm S}_{21}$, instead, we replace the Taylor-expansions
$(f^{\rm S}_{0},f_{1}^{\rm S})$ in Eq.~\eqref{eq:hatfS} by 
$(\overline{f^{\rm S}_{0}},\overline{f_{1}^{\rm S}})$, because $X_{12}=0$ when $\nu=1/4$ 
and thus the direct inverse resummation of $\hat{f}_{21}^{\rm S}$ is
singular at $\nu=1/4$. To validate this procedure we applied it to the 
test-particle limit of Eq.~\eqref{eq:hatfS} (i.e., without the $\hat{a}^{3}$ term) 
finding again a remarkable agreement with $f_{21}^{\rm Num}$ 
(see red-dashed line in Fig.~\ref{fig:fig3}) all over.

%------------------
Figures~\ref{fig:A22}-\ref{fig:A21} show that: (i) for $(2,2)$ mode the effect of iResum 
(both at NLO and NNLO level) is to reduce NR/EOB gap at merger 
as well as the difference between NLO and NNLO approximants; surprisingly, even with 
iResum, the NR/EOB difference {\it is not} smaller at NNLO than at NLO 
(especially for $\ha_{1}=\ha_{2}=+0.98$). This is due to the excessive growth 
of $\overline{\hat{\rho}_{22}^{\rm S}}$ that is not compensated by the 
(decreasing) $\rho_{22}^{\rm orb}$. Interestingly, the  problem can be solved 
proceeding similarly to $\hat{f}^{\rm S}_{21}$,  i.e., writing Eq.~\eqref{eq:trho22} 
as $\hat{\rho}_{22}^{\rm S}=(1 -\frac{\ha_{0}}{2}x^{3/2}+\dots) -\frac{1}{6}X_{12}\tilde{a}_{12} x^{3/2}(1-\frac{221}{42}\ha_{0}x^{3/2}+\dots)$
and then iResumming separately each $1+\dots$ piece. The result 
(magenta lines in Fig.~\ref{fig:A22}), shows perfect consistency between NLO and NNLO 
and suggests that the NNLO information has actually little impact;
(ii) Fig.~\ref{fig:A21} tells a similar story for the $(2,1)$ mode (for a different binary though, 
since $h_{21}=0$ for $m_{1}=m_{2}$ and $\ha_{1}=\ha_{2}$), with the iResum NLO 
well consistent within the NR waveform error (typically, of a few percents) 
and essentially comparable to (any of) the NNLO amplitudes.

\section{Conclusions}
Our results (both for $\nu\neq0$ and $\nu=0$) indicate that
iResum waveforms may be better analytical choice for EOBNR models, with little NR-tuned additional modifications (e.g., by fixing the
NQC factors~\cite{Nagar:2015xqa}) needed to obtain an excellent
EOB/NR amplitude agreement at merger. Our approach looks particularly
promising (and in fact needed) to easily improve subdominant multipoles
(as the $(2,1)$, discussed in detail) that are currently (mostly) missing
in EOBNR waveform models. The resummation of the other subdominant multipoles,
as well as their consistent implementation in {\tt SEOBNR\_ihes}, so as to
improve both the waveform and the radiation reaction, will be discussed in future work

\acknowledgments
We are very grateful to S.~Hughes for providing us with the numerical fluxes used
to compute $(\rho_{\lm}^{\rm Num},f_{21}^{\rm Num})$. We warmly thank 
A.~Boh\'e, G.~Faye and S.~Marsat for computing for us the multipolar decomposition of 
their PN total fluxes. A.~N. is very much indebted to: D.~Hilditch for discussions that occurred
long ago and that eventually helped to shape a concept; to E.~Harms to compute and 
test, at a very early stage of this work, the behavior of the standard $\rho_{22}$ up to 
14PN; and T.~Damour for discussions and timely constructive criticisms.
A.~S. thanks IHES for hospitality at various stages of development of this work.
This work was supported in part by the European Research Council under the European
Union's Seventh Framework Programme (FP7/2007-2013)/ERC grant agreement no. 304978.

\bibliography{refs20160928}

\begin{thebibliography}{31}
\expandafter\ifx\csname natexlab\endcsname\relax\def\natexlab#1{#1}\fi
\expandafter\ifx\csname bibnamefont\endcsname\relax
  \def\bibnamefont#1{#1}\fi
\expandafter\ifx\csname bibfnamefont\endcsname\relax
  \def\bibfnamefont#1{#1}\fi
\expandafter\ifx\csname citenamefont\endcsname\relax
  \def\citenamefont#1{#1}\fi
\expandafter\ifx\csname url\endcsname\relax
  \def\url#1{\texttt{#1}}\fi
\expandafter\ifx\csname urlprefix\endcsname\relax\def\urlprefix{URL }\fi
\providecommand{\bibinfo}[2]{#2}
\providecommand{\eprint}[2][]{\url{#2}}

\bibitem[{\citenamefont{Abbott et~al.}(2016{\natexlab{a}})}]{Abbott:2016blz}
\bibinfo{author}{\bibfnamefont{B.~P.} \bibnamefont{Abbott}}
  \bibnamefont{et~al.} (\bibinfo{collaboration}{Virgo, LIGO Scientific}),
  \bibinfo{journal}{Phys. Rev. Lett.} \textbf{\bibinfo{volume}{116}},
  \bibinfo{pages}{061102} (\bibinfo{year}{2016}{\natexlab{a}}),
  \eprint{1602.03837}.

\bibitem[{\citenamefont{Abbott et~al.}(2016{\natexlab{b}})}]{Abbott:2016nmj}
\bibinfo{author}{\bibfnamefont{B.~P.} \bibnamefont{Abbott}}
  \bibnamefont{et~al.} (\bibinfo{collaboration}{Virgo, LIGO Scientific}),
  \bibinfo{journal}{Phys. Rev. Lett.} \textbf{\bibinfo{volume}{116}},
  \bibinfo{pages}{241103} (\bibinfo{year}{2016}{\natexlab{b}}),
  \eprint{1606.04855}.

\bibitem[{\citenamefont{Taracchini
  et~al.}(2014{\natexlab{a}})\citenamefont{Taracchini, Buonanno, Pan, Hinderer,
  Boyle et~al.}}]{Taracchini:2013rva}
\bibinfo{author}{\bibfnamefont{A.}~\bibnamefont{Taracchini}},
  \bibinfo{author}{\bibfnamefont{A.}~\bibnamefont{Buonanno}},
  \bibinfo{author}{\bibfnamefont{Y.}~\bibnamefont{Pan}},
  \bibinfo{author}{\bibfnamefont{T.}~\bibnamefont{Hinderer}},
  \bibinfo{author}{\bibfnamefont{M.}~\bibnamefont{Boyle}},
  \bibnamefont{et~al.}, \bibinfo{journal}{Phys.Rev.}
  \textbf{\bibinfo{volume}{D89}}, \bibinfo{pages}{061502}
  (\bibinfo{year}{2014}{\natexlab{a}}), \eprint{1311.2544}.

\bibitem[{\citenamefont{Husa et~al.}(2016)\citenamefont{Husa, Khan, Hannam,
  Puerrer, Ohme, Jiménez~Forteza, and Boh\'e}}]{Husa:2015iqa}
\bibinfo{author}{\bibfnamefont{S.}~\bibnamefont{Husa}},
  \bibinfo{author}{\bibfnamefont{S.}~\bibnamefont{Khan}},
  \bibinfo{author}{\bibfnamefont{M.}~\bibnamefont{Hannam}},
  \bibinfo{author}{\bibfnamefont{M.}~\bibnamefont{Puerrer}},
  \bibinfo{author}{\bibfnamefont{F.}~\bibnamefont{Ohme}},
  \bibinfo{author}{\bibfnamefont{X.}~\bibnamefont{Jiménez~Forteza}},
  \bibnamefont{and} \bibinfo{author}{\bibfnamefont{A.}~\bibnamefont{Boh\'e}},
  \bibinfo{journal}{Phys. Rev.} \textbf{\bibinfo{volume}{D93}},
  \bibinfo{pages}{044006} (\bibinfo{year}{2016}), \eprint{1508.07250}.

\bibitem[{\citenamefont{Khan et~al.}(2016)\citenamefont{Khan, Husa, Hannam,
  Ohme, Puerrer, Jiménez~Forteza, and Boh\'e}}]{Khan:2015jqa}
\bibinfo{author}{\bibfnamefont{S.}~\bibnamefont{Khan}},
  \bibinfo{author}{\bibfnamefont{S.}~\bibnamefont{Husa}},
  \bibinfo{author}{\bibfnamefont{M.}~\bibnamefont{Hannam}},
  \bibinfo{author}{\bibfnamefont{F.}~\bibnamefont{Ohme}},
  \bibinfo{author}{\bibfnamefont{M.}~\bibnamefont{Puerrer}},
  \bibinfo{author}{\bibfnamefont{X.}~\bibnamefont{Jiménez~Forteza}},
  \bibnamefont{and} \bibinfo{author}{\bibfnamefont{A.}~\bibnamefont{Boh\'e}},
  \bibinfo{journal}{Phys. Rev.} \textbf{\bibinfo{volume}{D93}},
  \bibinfo{pages}{044007} (\bibinfo{year}{2016}), \eprint{1508.07253}.

\bibitem[{\citenamefont{Damour and Nagar}(2014)}]{Damour:2014sva}
\bibinfo{author}{\bibfnamefont{T.}~\bibnamefont{Damour}} \bibnamefont{and}
  \bibinfo{author}{\bibfnamefont{A.}~\bibnamefont{Nagar}},
  \bibinfo{journal}{Phys.Rev.} \textbf{\bibinfo{volume}{D90}},
  \bibinfo{pages}{044018} (\bibinfo{year}{2014}), \eprint{1406.6913}.

\bibitem[{\citenamefont{Szilágyi et~al.}(2015)\citenamefont{Szilágyi,
  Blackman, Buonanno, Taracchini, Pfeiffer, Scheel, Chu, Kidder, and
  Pan}}]{Szilagyi:2015rwa}
\bibinfo{author}{\bibfnamefont{B.}~\bibnamefont{Szilágyi}},
  \bibinfo{author}{\bibfnamefont{J.}~\bibnamefont{Blackman}},
  \bibinfo{author}{\bibfnamefont{A.}~\bibnamefont{Buonanno}},
  \bibinfo{author}{\bibfnamefont{A.}~\bibnamefont{Taracchini}},
  \bibinfo{author}{\bibfnamefont{H.~P.} \bibnamefont{Pfeiffer}},
  \bibinfo{author}{\bibfnamefont{M.~A.} \bibnamefont{Scheel}},
  \bibinfo{author}{\bibfnamefont{T.}~\bibnamefont{Chu}},
  \bibinfo{author}{\bibfnamefont{L.~E.} \bibnamefont{Kidder}},
  \bibnamefont{and} \bibinfo{author}{\bibfnamefont{Y.}~\bibnamefont{Pan}},
  \bibinfo{journal}{Phys. Rev. Lett.} \textbf{\bibinfo{volume}{115}},
  \bibinfo{pages}{031102} (\bibinfo{year}{2015}), \eprint{1502.04953}.

\bibitem[{\citenamefont{Nagar et~al.}(2016)\citenamefont{Nagar, Damour,
  Reisswig, and Pollney}}]{Nagar:2015xqa}
\bibinfo{author}{\bibfnamefont{A.}~\bibnamefont{Nagar}},
  \bibinfo{author}{\bibfnamefont{T.}~\bibnamefont{Damour}},
  \bibinfo{author}{\bibfnamefont{C.}~\bibnamefont{Reisswig}}, \bibnamefont{and}
  \bibinfo{author}{\bibfnamefont{D.}~\bibnamefont{Pollney}},
  \bibinfo{journal}{Phys. Rev.} \textbf{\bibinfo{volume}{D93}},
  \bibinfo{pages}{044046} (\bibinfo{year}{2016}), \eprint{1506.08457}.

\bibitem[{\citenamefont{Abbott
  et~al.}(2016{\natexlab{c}})}]{TheLIGOScientific:2016wfe}
\bibinfo{author}{\bibfnamefont{B.~P.} \bibnamefont{Abbott}}
  \bibnamefont{et~al.} (\bibinfo{collaboration}{Virgo, LIGO Scientific}),
  \bibinfo{journal}{Phys. Rev. Lett.} \textbf{\bibinfo{volume}{116}},
  \bibinfo{pages}{241102} (\bibinfo{year}{2016}{\natexlab{c}}),
  \eprint{1602.03840}.

\bibitem[{\citenamefont{Damour and Nagar}(2007)}]{Damour:2007xr}
\bibinfo{author}{\bibfnamefont{T.}~\bibnamefont{Damour}} \bibnamefont{and}
  \bibinfo{author}{\bibfnamefont{A.}~\bibnamefont{Nagar}},
  \bibinfo{journal}{Phys. Rev.} \textbf{\bibinfo{volume}{D76}},
  \bibinfo{pages}{064028} (\bibinfo{year}{2007}), \eprint{0705.2519}.

\bibitem[{\citenamefont{Damour et~al.}(2009)\citenamefont{Damour, Iyer, and
  Nagar}}]{Damour:2008gu}
\bibinfo{author}{\bibfnamefont{T.}~\bibnamefont{Damour}},
  \bibinfo{author}{\bibfnamefont{B.~R.} \bibnamefont{Iyer}}, \bibnamefont{and}
  \bibinfo{author}{\bibfnamefont{A.}~\bibnamefont{Nagar}},
  \bibinfo{journal}{Phys. Rev.} \textbf{\bibinfo{volume}{D79}},
  \bibinfo{pages}{064004} (\bibinfo{year}{2009}).

\bibitem[{\citenamefont{Pan et~al.}(2011)\citenamefont{Pan, Buonanno, Fujita,
  Racine, and Tagoshi}}]{Pan:2010hz}
\bibinfo{author}{\bibfnamefont{Y.}~\bibnamefont{Pan}},
  \bibinfo{author}{\bibfnamefont{A.}~\bibnamefont{Buonanno}},
  \bibinfo{author}{\bibfnamefont{R.}~\bibnamefont{Fujita}},
  \bibinfo{author}{\bibfnamefont{E.}~\bibnamefont{Racine}}, \bibnamefont{and}
  \bibinfo{author}{\bibfnamefont{H.}~\bibnamefont{Tagoshi}},
  \bibinfo{journal}{Phys.Rev.} \textbf{\bibinfo{volume}{D83}},
  \bibinfo{pages}{064003} (\bibinfo{year}{2011}), \eprint{1006.0431}.

\bibitem[{\citenamefont{Damour et~al.}(1998)\citenamefont{Damour, Iyer, and
  Sathyaprakash}}]{Damour:1997ub}
\bibinfo{author}{\bibfnamefont{T.}~\bibnamefont{Damour}},
  \bibinfo{author}{\bibfnamefont{B.~R.} \bibnamefont{Iyer}}, \bibnamefont{and}
  \bibinfo{author}{\bibfnamefont{B.~S.} \bibnamefont{Sathyaprakash}},
  \bibinfo{journal}{Phys. Rev.} \textbf{\bibinfo{volume}{D57}},
  \bibinfo{pages}{885} (\bibinfo{year}{1998}), \eprint{gr-qc/9708034}.

\bibitem[{\citenamefont{Blanchet}(2014)}]{Blanchet:2013haa}
\bibinfo{author}{\bibfnamefont{L.}~\bibnamefont{Blanchet}},
  \bibinfo{journal}{Living Rev.Rel.} \textbf{\bibinfo{volume}{17}},
  \bibinfo{pages}{2} (\bibinfo{year}{2014}), \eprint{1310.1528}.

\bibitem[{\citenamefont{Nagar and Akcay}(2012)}]{Nagar:2011aa}
\bibinfo{author}{\bibfnamefont{A.}~\bibnamefont{Nagar}} \bibnamefont{and}
  \bibinfo{author}{\bibfnamefont{S.}~\bibnamefont{Akcay}},
  \bibinfo{journal}{Phys.Rev.} \textbf{\bibinfo{volume}{D85}},
  \bibinfo{pages}{044025} (\bibinfo{year}{2012}), \eprint{1112.2840}.

\bibitem[{\citenamefont{Taracchini et~al.}(2013)\citenamefont{Taracchini,
  Buonanno, Hughes, and Khanna}}]{Taracchini:2013wfa}
\bibinfo{author}{\bibfnamefont{A.}~\bibnamefont{Taracchini}},
  \bibinfo{author}{\bibfnamefont{A.}~\bibnamefont{Buonanno}},
  \bibinfo{author}{\bibfnamefont{S.~A.} \bibnamefont{Hughes}},
  \bibnamefont{and} \bibinfo{author}{\bibfnamefont{G.}~\bibnamefont{Khanna}},
  \bibinfo{journal}{Phys.Rev.} \textbf{\bibinfo{volume}{D88}},
  \bibinfo{pages}{044001} (\bibinfo{year}{2013}), \eprint{1305.2184}.

\bibitem[{\citenamefont{Taracchini et~al.}(2012)\citenamefont{Taracchini, Pan,
  Buonanno, Barausse, Boyle et~al.}}]{Taracchini:2012ig}
\bibinfo{author}{\bibfnamefont{A.}~\bibnamefont{Taracchini}},
  \bibinfo{author}{\bibfnamefont{Y.}~\bibnamefont{Pan}},
  \bibinfo{author}{\bibfnamefont{A.}~\bibnamefont{Buonanno}},
  \bibinfo{author}{\bibfnamefont{E.}~\bibnamefont{Barausse}},
  \bibinfo{author}{\bibfnamefont{M.}~\bibnamefont{Boyle}},
  \bibnamefont{et~al.}, \bibinfo{journal}{Phys.Rev.}
  \textbf{\bibinfo{volume}{D86}}, \bibinfo{pages}{024011}
  (\bibinfo{year}{2012}), \eprint{1202.0790}.

\bibitem[{\citenamefont{Damour and Nagar}(2009)}]{Damour:2009kr}
\bibinfo{author}{\bibfnamefont{T.}~\bibnamefont{Damour}} \bibnamefont{and}
  \bibinfo{author}{\bibfnamefont{A.}~\bibnamefont{Nagar}},
  \bibinfo{journal}{Phys. Rev.} \textbf{\bibinfo{volume}{D79}},
  \bibinfo{pages}{081503} (\bibinfo{year}{2009}).

\bibitem[{\citenamefont{Marsat et~al.}(2013)\citenamefont{Marsat, Blanchet,
  Boh\'e, and Faye}}]{Marsat:2013wwa}
\bibinfo{author}{\bibfnamefont{S.}~\bibnamefont{Marsat}},
  \bibinfo{author}{\bibfnamefont{L.}~\bibnamefont{Blanchet}},
  \bibinfo{author}{\bibfnamefont{A.}~\bibnamefont{Boh\'e}}, \bibnamefont{and}
  \bibinfo{author}{\bibfnamefont{G.}~\bibnamefont{Faye}}
  (\bibinfo{year}{2013}), \eprint{1312.5375}.

\bibitem[{\citenamefont{Marsat}(2015)}]{Marsat:2014xea}
\bibinfo{author}{\bibfnamefont{S.}~\bibnamefont{Marsat}},
  \bibinfo{journal}{Class. Quant. Grav.} \textbf{\bibinfo{volume}{32}},
  \bibinfo{pages}{085008} (\bibinfo{year}{2015}), \eprint{1411.4118}.

\bibitem[{\citenamefont{Boh\'e et~al.}(2015)\citenamefont{Boh\'e, Faye, Marsat,
  and Porter}}]{Bohe:2015ana}
\bibinfo{author}{\bibfnamefont{A.}~\bibnamefont{Boh\'e}},
  \bibinfo{author}{\bibfnamefont{G.}~\bibnamefont{Faye}},
  \bibinfo{author}{\bibfnamefont{S.}~\bibnamefont{Marsat}}, \bibnamefont{and}
  \bibinfo{author}{\bibfnamefont{E.~K.} \bibnamefont{Porter}},
  \bibinfo{journal}{Class. Quant. Grav.} \textbf{\bibinfo{volume}{32}},
  \bibinfo{pages}{195010} (\bibinfo{year}{2015}), \eprint{1501.01529}.

\bibitem[{\citenamefont{Fujita}(2015)}]{Fujita:2014eta}
\bibinfo{author}{\bibfnamefont{R.}~\bibnamefont{Fujita}},
  \bibinfo{journal}{PTEP} \textbf{\bibinfo{volume}{2015}},
  \bibinfo{pages}{033E01} (\bibinfo{year}{2015}), \eprint{1412.5689}.

\bibitem[{\citenamefont{Shah}(2014)}]{Shah:2014tka}
\bibinfo{author}{\bibfnamefont{A.~G.} \bibnamefont{Shah}},
  \bibinfo{journal}{Phys. Rev.} \textbf{\bibinfo{volume}{D90}},
  \bibinfo{pages}{044025} (\bibinfo{year}{2014}), \eprint{1403.2697}.

\bibitem[{\citenamefont{Isoyama et~al.}(2013)\citenamefont{Isoyama, Fujita,
  Sago, Tagoshi, and Tanaka}}]{Isoyama:2012bx}
\bibinfo{author}{\bibfnamefont{S.}~\bibnamefont{Isoyama}},
  \bibinfo{author}{\bibfnamefont{R.}~\bibnamefont{Fujita}},
  \bibinfo{author}{\bibfnamefont{N.}~\bibnamefont{Sago}},
  \bibinfo{author}{\bibfnamefont{H.}~\bibnamefont{Tagoshi}}, \bibnamefont{and}
  \bibinfo{author}{\bibfnamefont{T.}~\bibnamefont{Tanaka}},
  \bibinfo{journal}{Phys. Rev.} \textbf{\bibinfo{volume}{D87}},
  \bibinfo{pages}{024010} (\bibinfo{year}{2013}), \eprint{1210.2569}.

\bibitem[{\citenamefont{Johnson-McDaniel}(2014)}]{Johnson-McDaniel:2014lia}
\bibinfo{author}{\bibfnamefont{N.~K.} \bibnamefont{Johnson-McDaniel}},
  \bibinfo{journal}{Phys. Rev.} \textbf{\bibinfo{volume}{D90}},
  \bibinfo{pages}{024043} (\bibinfo{year}{2014}), \eprint{1405.1572}.

\bibitem[{\citenamefont{Taracchini
  et~al.}(2014{\natexlab{b}})\citenamefont{Taracchini, Buonanno, Khanna, and
  Hughes}}]{Taracchini:2014zpa}
\bibinfo{author}{\bibfnamefont{A.}~\bibnamefont{Taracchini}},
  \bibinfo{author}{\bibfnamefont{A.}~\bibnamefont{Buonanno}},
  \bibinfo{author}{\bibfnamefont{G.}~\bibnamefont{Khanna}}, \bibnamefont{and}
  \bibinfo{author}{\bibfnamefont{S.~A.} \bibnamefont{Hughes}}
  (\bibinfo{year}{2014}{\natexlab{b}}), \eprint{1404.1819}.

\bibitem[{\citenamefont{Harms et~al.}(2014)\citenamefont{Harms, Bernuzzi,
  Nagar, and Zenginoglu}}]{Harms:2014dqa}
\bibinfo{author}{\bibfnamefont{E.}~\bibnamefont{Harms}},
  \bibinfo{author}{\bibfnamefont{S.}~\bibnamefont{Bernuzzi}},
  \bibinfo{author}{\bibfnamefont{A.}~\bibnamefont{Nagar}}, \bibnamefont{and}
  \bibinfo{author}{\bibfnamefont{A.}~\bibnamefont{Zenginoglu}}
  (\bibinfo{year}{2014}), \eprint{1406.5983}.

\bibitem[{\citenamefont{Nagar et~al.}(2014)\citenamefont{Nagar, Harms,
  Bernuzzi, and Zenginoğlu}}]{Nagar:2014kha}
\bibinfo{author}{\bibfnamefont{A.}~\bibnamefont{Nagar}},
  \bibinfo{author}{\bibfnamefont{E.}~\bibnamefont{Harms}},
  \bibinfo{author}{\bibfnamefont{S.}~\bibnamefont{Bernuzzi}}, \bibnamefont{and}
  \bibinfo{author}{\bibfnamefont{A.}~\bibnamefont{Zenginoğlu}},
  \bibinfo{journal}{Phys. Rev.} \textbf{\bibinfo{volume}{D90}},
  \bibinfo{pages}{124086} (\bibinfo{year}{2014}), \eprint{1407.5033}.

\bibitem[{SXS()}]{SXS:catalog}
\bibinfo{howpublished}{\url{http://www.black-holes.org/waveforms}}.

\bibitem[{\citenamefont{Harms et~al.}(2016)\citenamefont{Harms,
  Lukes-Gerakopoulos, Bernuzzi, and Nagar}}]{Harms:2015ixa}
\bibinfo{author}{\bibfnamefont{E.}~\bibnamefont{Harms}},
  \bibinfo{author}{\bibfnamefont{G.}~\bibnamefont{Lukes-Gerakopoulos}},
  \bibinfo{author}{\bibfnamefont{S.}~\bibnamefont{Bernuzzi}}, \bibnamefont{and}
  \bibinfo{author}{\bibfnamefont{A.}~\bibnamefont{Nagar}},
  \bibinfo{journal}{Phys. Rev.} \textbf{\bibinfo{volume}{D93}},
  \bibinfo{pages}{044015} (\bibinfo{year}{2016}), \eprint{1510.05548}.

\bibitem[{\citenamefont{Gralla et~al.}(2016)\citenamefont{Gralla, Hughes, and
  Warburton}}]{Gralla:2016qfw}
\bibinfo{author}{\bibfnamefont{S.~E.} \bibnamefont{Gralla}},
  \bibinfo{author}{\bibfnamefont{S.~A.} \bibnamefont{Hughes}},
  \bibnamefont{and}
  \bibinfo{author}{\bibfnamefont{N.}~\bibnamefont{Warburton}},
  \bibinfo{journal}{Class. Quant. Grav.} \textbf{\bibinfo{volume}{33}},
  \bibinfo{pages}{155002} (\bibinfo{year}{2016}), \eprint{1603.01221}.

\end{thebibliography}

\end{document}